\documentclass[onelcolumn]{aastex631}

\shorttitle{Wave Reflection in the Solar Atmosphere}
\shortauthors{Chaturmutha et al.}

\usepackage{float}
\usepackage{graphicx}
\graphicspath{{Figures}{.}}
\usepackage{mathtools}
\begin{document}

\title{Wave Reflection in the Solar Atmosphere}

\correspondingauthor{Varun Chaturmutha}
\email{vchaturmutha@gsu.edu}

\author[0000-0003-3220-289X]{Varun Chaturmutha}
\affiliation{Georgia State University, 25 Park Place, Atlanta, GA 30303, USA}

\author[0000-0001-5777-9121]{Bernhard Fleck}
\affil{ESA Science and Operations Department, c/o NASA/GSFC Code 671, Greenbelt, MD 20071, USA}

\author[0000-0002-9580-5615]{Stuart Jefferies}
\affiliation{Georgia State University, 25 Park Place, Atlanta, GA 30303, USA}
\affiliation{University of Hawaii, 34, Ohia Ku Street, Pukalani, HI 96768, USA}



\begin{abstract}

We present evidence supporting wave reflection in the lower solar chromosphere based on helioseismic analysis of multi-height Doppler data. This evidence is derived through a wave propagation model that incorporates both upward- and downward-traveling (reflected) waves. Moreover, we find that the height of the reflecting region varies with magnetic field strengths in a way that suggests a connection with the the plasma $\beta\sim1$ region. We measure an effective reflection coefficient of $13\,\%$ in a magnetically quiet region of the Sun.

\end{abstract}



\section{Introduction}
\label{sec:intro}

Acoustic waves are generated at the base of the photosphere through turbulent motions in the convection zone and they travel isotropically. While some waves travel upward, others traveling downward undergo refraction in the interior. This refraction is due to the increasing sound speed with depth, gradually redirecting them upward toward the top of the convection zone -- the convection zone-photosphere (CZP) boundary ($\sim0\,$Mm).

Among these waves that reach the CZP boundary, and the originally upward emitted waves, only those with frequencies above the acoustic cutoff frequency travel as propagating waves into the photosphere, whereas lower-frequency waves become evanescent in the photosphere. Observing the phase travel times of the propagating acoustic waves through Doppler shifts of spectral lines formed at different heights in the solar atmosphere has the potential to provide valuable insights into the properties of the solar atmosphere. 

Previous works have demonstrated that waves traveling beyond the CZP boundary undergo further reflections in the upper atmosphere. \textcolor{black}{\cite{1976SoPh...49..231M} show that magneto-acoustic waves traveling upward undergo reflection at the chromosphere-corona transition region (CCTR; $\sim2\,$Mm) due to the steep gradient of Alfvén Speed.} There have also been suggestions for an additional reflective layer in the lower chromosphere ($\sim 1\,$Mm). \cite{1989A&A...224..245F} referred this as the ``magic height'' in the lower chromosphere beyond which the phase lags do not propagate higher up in the atmosphere. This region could correspond to the reflective layer reported in our study. Additional support for the presence of a reflecting layer in the lower chromosphere comes from the modulation of time-distance diagrams for acoustic waves with frequencies $> 5\,$mHz \citep{1997ApJ...485L..49J}. Moreover, the existence of traveling waves that are reflected near the lower chromosphere may explain the observed power halos around active regions \citep{2007A&A...471..961M, 2009A&A...506L...5K, 2015ApJ...801...27R}.

\cite{2019ApJ...884L...8J} suggested the presence of a signature of downward propagating waves in the atmosphere, in the form of a ``wiggle” in the frequency variation of their 1-D phase difference spectra, that is not accounted for by a model that only considers upward propagating waves \citep{1972A&A....17..458S}. In this work, we build upon this suggestion and the work of \cite{1997ApJ...485L..49J}, and model the observed phase travel times using a low-Q Fabry–Pérot model where acoustic waves with frequencies above the acoustic cut-off frequency are partially reflected at the CZP boundary and a layer in the upper atmosphere. This model is analogous to the model used to explain asymmetries in the solar oscillation line profiles \citep{1993ApJ...410..829D}.

The outline of the paper is as follows. In Section \ref{sec:obsdata}, we describe the observational data sets and analysis techniques used to investigate wave propagation. Section \ref{sec:owm} discusses the upward propagating wave model that has been used for the past 40 years. \ref{sec:twm} introduces a multi-wave model incorporating both upward and downward propagating waves. \ref{sec:near-deg} brings to attention the near-degeneracy issue with aforementioned wave propagation models. Section \ref{sec:worrall} explores the phase travel time behaviors using simulations based on a rectangular barrier model in comparison with observations. Results are discussed in Section \ref{sec:res} and Section \ref{sec:conclusion} concludes.

\section{Observations}
\label{sec:obsdata}
For our analysis here, we utilized two sets of multi-height Doppler data from two instruments: the space-based Helioseismic and Magnetic Imager (HMI) aboard the Solar Dynamics Observatory (SDO; \citealp{2012SoPh..275..207S, 2012SoPh..275..229S}) and the Magneto-Optical filters at Two Heights II (MOTH; \citealp{2018IAUS..335..335F}) instrument located at the South Pole. The first data set comprises the HMI Fe -- Fe Doppler pair, which was derived from the first and second Fourier coefficients \textcolor{black}{\citep{2012SoPh..278..217C}} of the Fe I $617.3\,$nm spectral line and was recorded on 2010 August 24 from 00:00:00 to 11:00:00 UT. This data set, hereafter referred to as the 2010 HMI Fe -- Fe data set, probes the lower photosphere ($\sim140-200\,$km; \textcolor{black}{\citealp{2014SoPh..289.3457N}}) with a separation of approximately $50\,$km between the Doppler pair formation heights. The second data set comprises the HMI Fe and MOTH Na $D_1$ ($589\,$nm) Doppler data that were acquired on 2017 January 21 from 07:01:30 to 18:01:30 UT. This data set, hereafter referred to as the 2017 MOTH Na -- HMI Fe data set, probes the upper photosphere/lower chromosphere at approximately $650\,$km using Na $D_1$ lines and the lower photosphere at approximately $140\,$km using the standard HMI Doppler velocities. Each Doppler data sets is 11 hours long with a cadence of 45\,s (the cadence of MOTH data was integrated from $5\,$s to $45\,$s to match the HMI cadence). Moreover, the HMI and MOTH data sets were binned spatially to yield an effective pixel size of 1.27 arcsec.  

The crosspower spectra are calculated as $CP(\omega) = F_{V_1}(\omega) \cdot F_{V_2}(\omega)^\star$, where $\omega$ is the cyclic frequency and $F$ is a Fourier transform of the measured velocity signals ($V_1$ and $V_2$) obtained from the Doppler pair in each of the data sets. The 2017 MOTH Na -- HMI Fe data sets contain spatial pixels $(x,y)$ within a region of $1060$\,arcsec\,$\times\,1060\,$arcsec near the center of the solar disk. For the 2010 HMI Fe -- Fe data set, we focused on a smaller region measuring $518$\,arcsec\,$\times\,518\,$arcsec, which corresponds to the central $1024\times1024$ pixels of the full $4096\times4096$ pixels resolution HMI Dopplergrams. \textcolor{black}{These are the same data sets as used in \cite{2019ApJ...884L...8J} where they also provide the magnetograms and acoustic power maps for the regions under study.}

\subsection{Observed Phase Travel Time Measurements}
\label{sec:obs-ph}

The phase travel time ($\Delta T_p$) between two heights can be calculated from the cross-power ($CP$) between the Doppler velocity signals at the two heights using the following expression,
\begin{equation}
    \label{eq:tt}
    \Delta T_p(\omega) = \frac{\Delta \Phi(\omega)}{\omega}.
\end{equation}
where $\Delta \Phi$ is the phase difference between the two heights, given by 
\begin{equation}
\label{eq:cp}
\Delta\Phi(\omega)=\arg(CP(\omega))=\arctan\left(\frac{I(\omega)}{R(\omega)}\right).
\end{equation}

Here, $\operatorname{arg(z)}$ represents the argument of a complex number ($z$) defined as $\operatorname{arg}(z) = \arctan\left(\frac{\operatorname{Imaginiary}(z)}{\operatorname{Real}(z)}\right)$, while $R$ and $I$ denote the real and imaginary parts of the cross power ($CP$). The standard deviation in the phase travel time measurements is given by:

\begin{equation}
    \label{eq:err}
    \sigma_{\Delta T_p}(\omega) = \pm \frac{\sqrt{\left(\sigma_{I}(\omega)\,/\,\overline I(\omega)\right)^2 + \left(\sigma_{R}(\omega)\,/\,\overline R(\omega)\right)^2}}{\overline I(\omega)^2 + \overline R(\omega)^2} \cdot \frac{1}{\omega},
\end{equation}

where $\sigma_{R}(\omega)$ and  $\sigma_{I}(\omega)$ are the standard deviations in the real and imaginary parts of the cross-power respectively along the spatial $(x,y)$ axes and $\overline I(\omega)$ and $\overline R(\omega)$ are the average imaginary and real parts of the cross-power respectively along the spatial $(x,y)$ axes. Here we use 3$\sigma_{\Delta T_p}$ as the standard error in our phase travel time measurements to include 99.7\% of the data.

The observed phase travel times measurements are shown in Figure \ref{fig:tt-disp} (a) and (b) using blue filled circles along with  vertical errors bars. Panels (a) and (b) correspond to the phase travel time measurements from the 2010 HMI Fe -- Fe and 2017 MOTH Na -- HMI Fe data sets respectively for the quiet-Sun low magnetic field region ($|B|<30\,$G). Figure \ref{fig:tt-disp3} shows a comparison of the phase travel time measurements for different magnetic field strengths in the 2010 HMI Fe -- Fe and 2017 MOTH Na -- HMI Fe data sets in panels (a) and (b) respectively: low magnetic field ($|B|<30\,$G; blue plus sign), intermediate magnetic field ($30\,$G$\,<|B|<100\,$G; green filled circles), and strong magnetic field ($|B|>100\,$G; red crosses).

\section{Wave Propagation Models}
\label{sec:wpm}

\subsection{One-Wave Model}
\label{sec:owm}

The observed phase travel time can be modeled using the dispersion relation for acoustic-gravity waves assuming an isothermal stratified atmosphere with a constant radiative cooling time \citep{1972A&A....17..458S}.
\textcolor{black}{The variables in the model are the height difference between the observing levels $\left(\Delta z\right)$, the 
the radiative cooling time $\left(\tau_r \right)$, sound speed  $\left(c_{s}\right)$, and horizontal wavenumber ($k_x$).}

To model the quiet-Sun phase travel time, we first define the vertical component of the velocity induced by an upward propagating plane-wave of angular frequency $\omega$ and angular wavenumber $k=\sqrt{k_x^2+k_z^2}$, where $k_x$ and $k_z$ are the horizontal and vertical components of the wavenumber respectively.
\begin{equation}
    \label{eq:v1}
    V(z) = V(0) e^{\kappa z} e^{
    i(\omega t - k_z z - k_x x)}
\end{equation}
where $V(0)$ is the amplitude of the wave at height $z\,=0\,\,$km (base of the photosphere) and $\kappa$ is related to the scale height of the atmosphere. 
The phase difference $\Delta\Phi$ between two heights, $z_1$ and $z_2$, is given by

\begin{equation}
    \label{eq:phi2}
    \Delta \Phi = \operatorname{arg}\{V(z_1)V^*(z_2)\},
\end{equation}

which when substituted with Equation \ref{eq:v1} gives,

\begin{equation}
    \label{eq:dphi}
    \Delta \Phi = k_z(z_2-z_1) = k_z\Delta z.
\end{equation}

The corresponding phase travel time for the One-Wave model can be calculated using Equation \ref{eq:tt},

\begin{equation}
\label{eq:tt2}
\Delta T(\omega) = k_z\Delta z/\omega.     
\end{equation}

Here, $k_z$ is a function of $\tau_r$, $c_s$, and $k_x$ given by \citep{1972A&A....17..458S}, 

\begin{equation}
    \label{eq:kz}
    k_z = \pm \left[ \frac{a}{2} + \frac{\sqrt{a^2+b^2}}{2}\right]^{\frac{1}{2}};
\end{equation},

\begin{equation}
    \label{eq:a}
    a = \frac{\omega^2-\omega_{ac}^2}{c_s^2} + k_x^2\left(\frac{N^2}{\omega^2} -1\right) - \frac{1}{1+\omega^2\tau_r^2} \frac{N^2}{\omega^2}\left(k_x^2 - \frac{\omega^4}{g^2}\right),
\end{equation}

\begin{equation}
    \label{eq:b}
    b = \frac{\omega\tau_r}{1+\omega^2\tau_r^2} \left(\frac{N^2}{\omega^2}\right)\left(k_x^2 - \frac{\omega^4}{g^2}\right).
\end{equation}

\textcolor{black}{The acoustic cutoff frequency $\left(\omega_{ac}\right)$ is a function of the sound speed, given as $\omega_{ac} = \gamma g /(2c_s)$} and the Brunt–Väisälä frequency ($N$) is given by the identity $N=(\gamma-1)^{\frac{1}{2}}g/c_s$. In the solar photosphere, the surface gravity $g=274\,$m/s$^2$ and the adiabatic constant is taken as $\gamma=5/3$.

\begin{figure*}
\gridline{
\fig{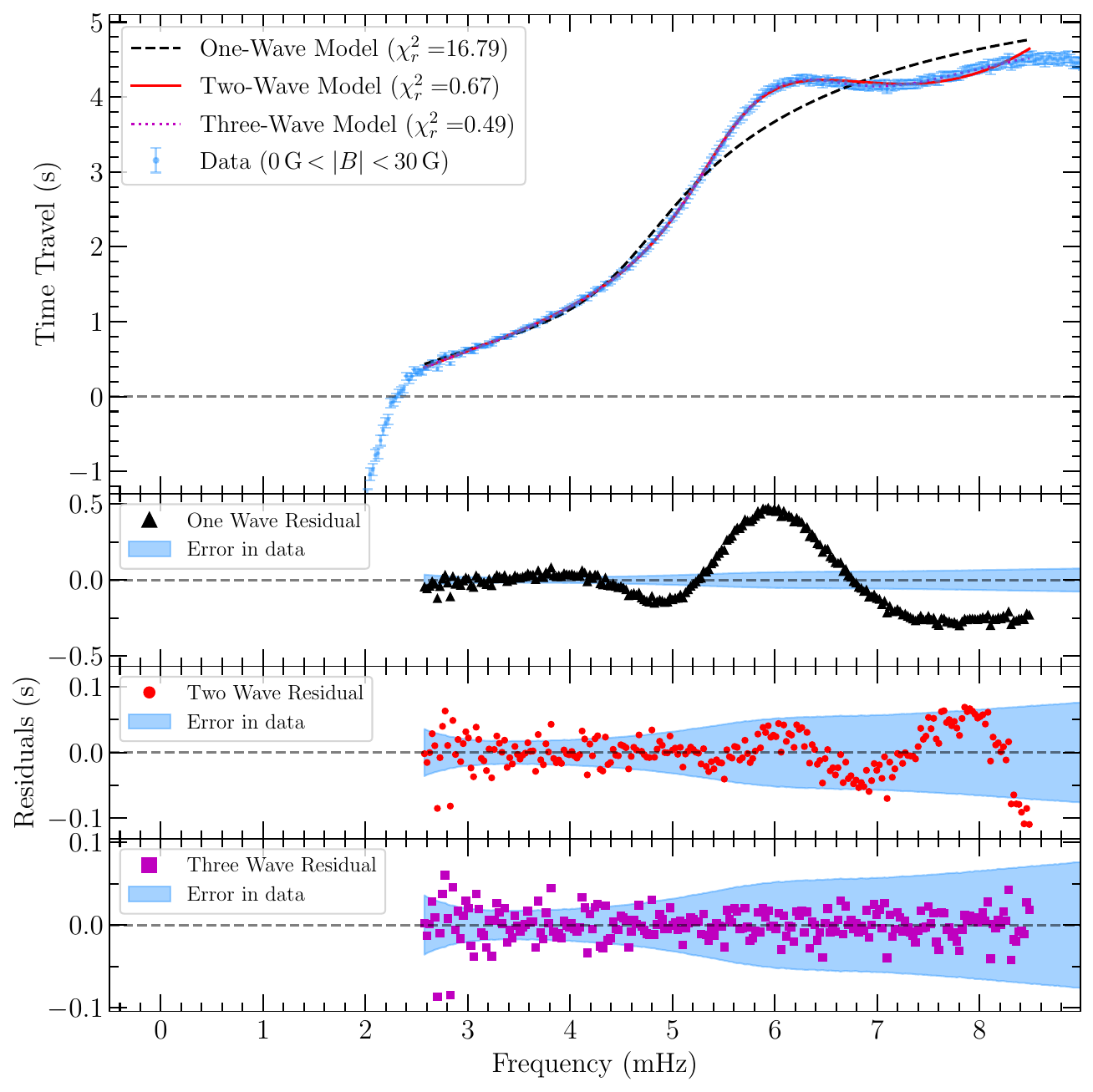}{0.5\textwidth}{(a) 2010 HMI Fe -- Fe\\ $|B|<30\,G$}
\fig{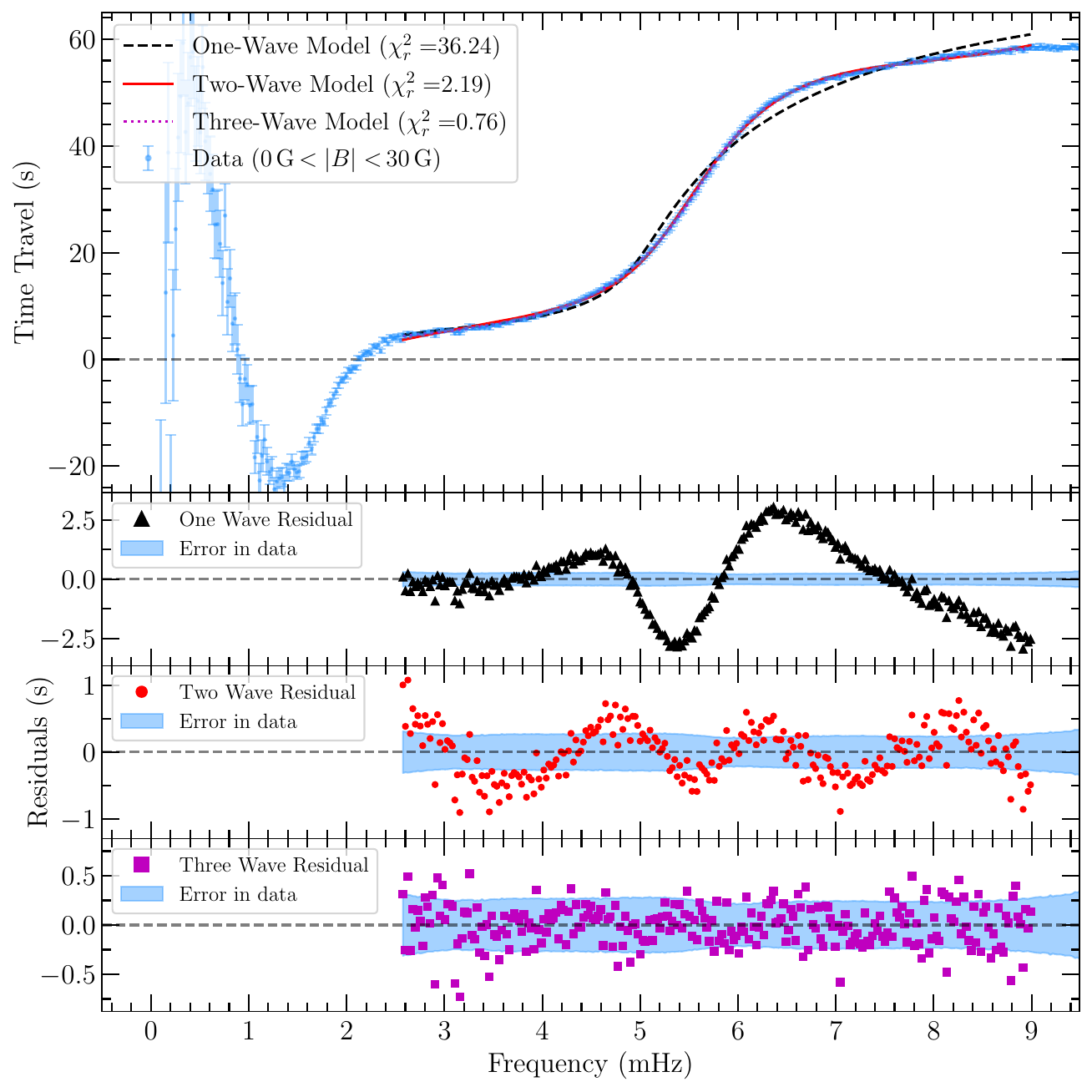}{0.5\textwidth}{(b) 2017 MOTH Na -- HMI Fe\\ $|B|<30\,G$}
         }
         \centering
\caption{(a) The observed phase travel time curve for the low magnetic field region in the 2010 HMI -- HMI (Fe -- Fe) in blue filled circles with vertical error bars in the observed phase travel time. The black dashed lines show the One-Wave fit and the black triangles plot the corresponding residuals in the panel below. The red solid line shows the Two-Wave fit and the red filled circles show the corresponding residuals in the panel below. Similarly, also plotted is the Three-Wave fit in purple dotted lines and residuals are shown in purple squares in the panel below. The error in the observed phase difference at each frequency is also plotted in the bottom panel using the shaded region of blue. \textit{Note: The phase travel time has been plotted here against frequency ($\nu$) which is given in terms of the cyclic frequency ($\omega$) as, $\nu=\omega/(2\pi)$}. (b) Identical to panel (a) but for the 2017 MOTH Na -- HMI Fe data set.
            \label{fig:tt-disp}
}
\end{figure*}

\begin{figure*}[t]
\gridline{
\fig{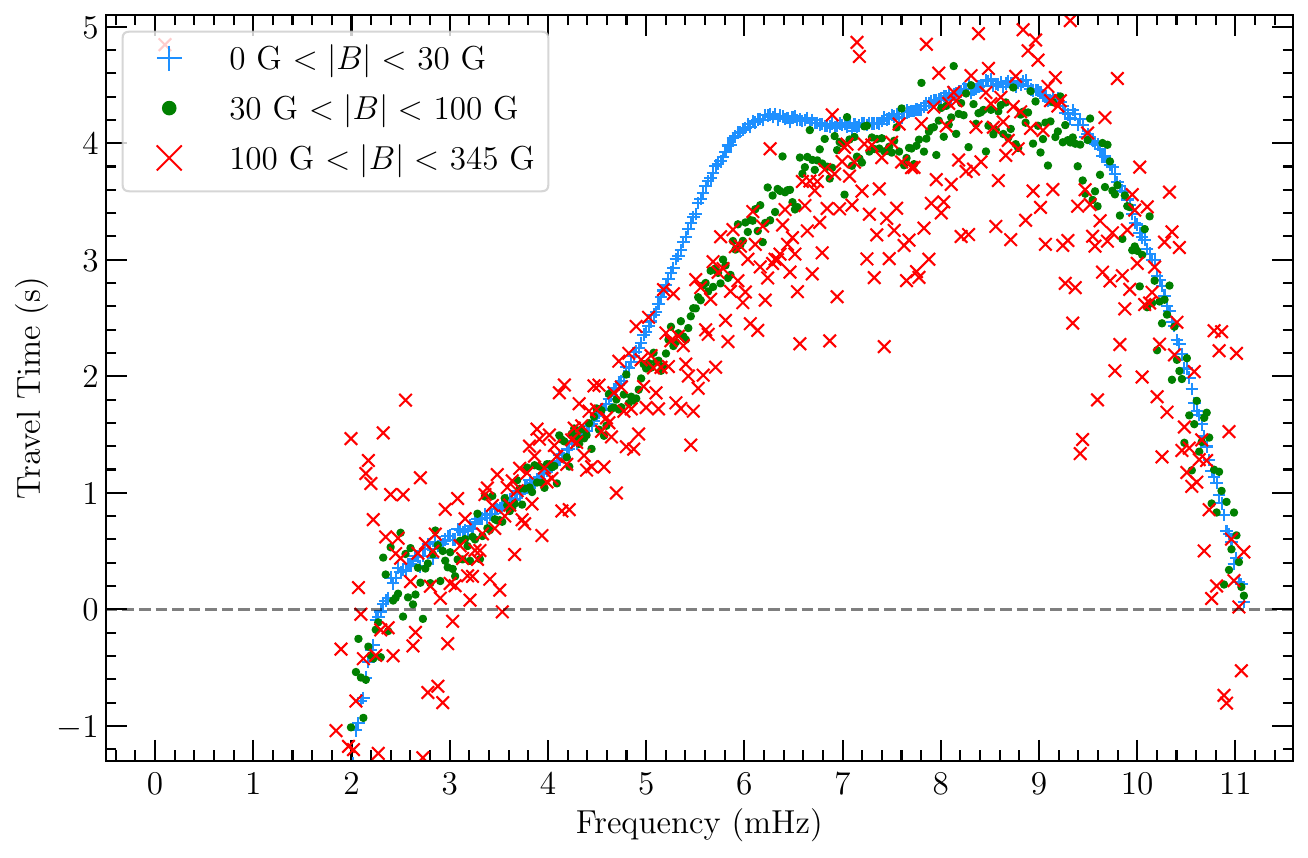}{0.5\textwidth}{(a) 2010 HMI Fe -- Fe}
\fig{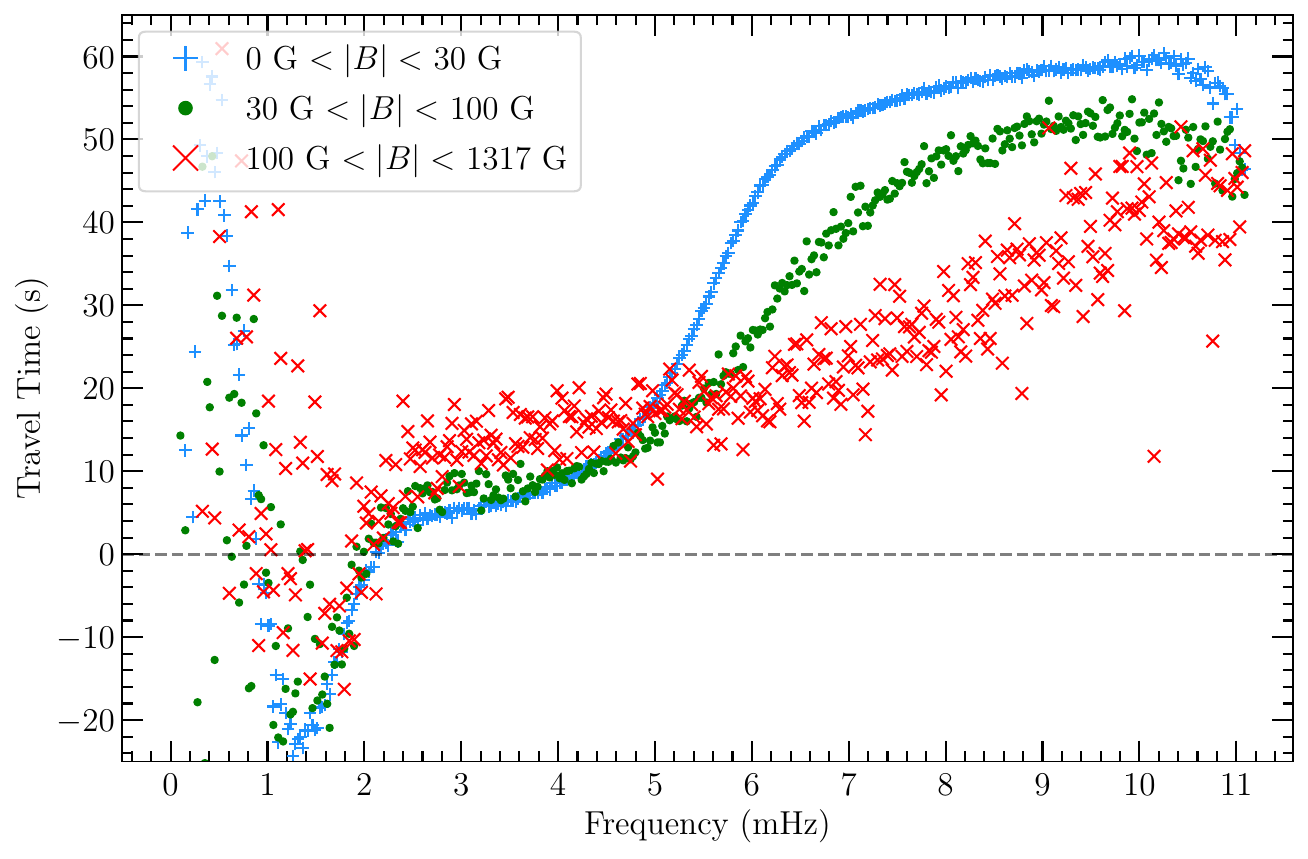}{0.5\textwidth}{(b) 2017 MOTH Na -- HMI Fe}
      }
\caption{(a) Comparison of the observed travel time curves for different magnetic field regions in the 2010 HMI Fe -- Fe data set where: the blue plus sign represents the low magnetic field region ($|B|<30\,G$), green filled circles represents the intermediate magnetic field region ($30<|B|<100\,G$), and the red crosses represents the high magnetic field region ($|B|>100\,G$). (b) Identical to (a) except that they are measurements from the 2017 MOTH Na -- HMI Fe data set. Both plots show variation in the observed travel time curves as the magnetic field strength changes. Most notably, the ``wiggle'' features corresponding to the wave reflection phenomenon disappears at strong magnetic field strengths. Moreover, the dispersion-free travel time at high frequencies in (b), the 2017 MOTH Na -- HMI Fe data set, shows a drop at stronger magnetic field strengths possibly because the reflective region drops below the upper observing level, thus reducing observed phase travel time at higher frequencies.
            \label{fig:tt-disp3}
}
\end{figure*}

The best One-Wave model fits, prepared using Equation \ref{eq:tt2}, to the quiet-Sun phase travel time curves are shown in Figure \ref{fig:tt-disp} using black dashed lines. The parameters corresponding to the One-Wave model fits are shown in Table \ref{tab:owm}. The uncertainties in these parameters are determined using a bootstrapping technique \citep{2010arXiv1012.3754A}. In this approach we generate 1000 samples each containing 100 randomly selected frequency points (from a pool of approximately 245 frequency points) and corresponding phase travel time values. The model is then fit to each of these samples. The reported parameter errors represent the standard deviations derived from the 1000 fitted parameter sets. The phase travel time curves are fit from \textcolor{black}{2.6\,mHz to 8.5\,mHz and 9\,mHz} for the 2010 MOTH Na -- HMI Fe and 2017 MOTH Na -- HMI data sets respectively. The lower limit corresponds to the frequency below which atmospheric gravity waves dominate and the higher frequency limit is chosen to avoid the roll-off of the phase travel time to zero at the higher frequencies. This behavior can be seen in Figure \ref{fig:tt-disp3} in panel (a) beyond 9\,mHz. Similar strong roll-offs can be seen in Figure 3 of \cite{2021RSPTA.37900170F} and Figure 5 of \cite{2022ApJ...933..109Z}. Panel (b) in Figure \ref{fig:tt-disp3} shows a roll-off starting around 10\,mHz.  Interestingly, strong roll-offs are observed for height separations of the measurement probes that are under 100\,km while more benign roll-offs are observed when the separations are around several 100\,km. A preliminary analysis with wave propagation models suggests that temporal under-sampling of the data plays a role, probably the dominant role, in the roll-off behavior and should be incorporated into the spectral modeling; especially as the spectral signal above the temporal cut-off frequency contains information on the ``wiggle'' in the travel time curve that provides a further constraint on the location of the reflecting region and the separation of the observing probes. Unfortunately, as the correction for under-sampling requires knowledge of the nature of the wave power at each observing height at the frequencies that are under-sampled, which we don’t have, further investigation is needed in how to overcome this lack of information. In the meantime, we restrict our fitting region to avoid the roll-off phenomenon.

The reduced chi-squared value for fits can be calculated as $\chi_r^2=\sum_\nu\frac{(O_\nu-E_\nu)^2}{\sigma_\nu^2}\cdot \frac{1}{(N-n)}$, where $O_\nu$ and $E_\nu$ are the observed and modeled phase travel time respectively, $\sigma_\nu$ is the standard error in the observed phase travel time, $N$ is the number of frequency points, and $n$ is the total number of free parameters in the model. The reduced chi-squared value for the 2010 Fe -- Fe and 2017 Na -- Fe data set is \textcolor{black}{$16.79$ \& $36.24$} respectively, that suggests a poor fit quality which is also evident by the large residuals (shown in black triangles) around $6\,$mHz in Figure \ref{fig:tt-disp} (a) and (b). The lack of flexibility to model the ``wiggle'' around $6\,$mHz was also noted in \cite{2019ApJ...884L...8J}. This indicates the presence of additional physical mechanisms that are not incorporated by the One-Wave model.

\subsection{Multi-Wave Model}
\label{sec:twm}

\begin{table*}
    \begin{tabular}{l|l|l|l|l|l}
    \hline
    Parameter & Parameter & \multicolumn{2}{c|}{Value $\pm$ Uncertainty} & \multicolumn{2}{c}{Bounds} \\
    \cline{3-6}
    Name & Notation (Unit) & 2010 Fe -- Fe & 2017 Na -- Fe & 2010 Fe -- Fe & 2017 Na -- Fe \\
    \hline
    \textcolor{black}{Height Difference} & $\Delta z$ (km) & $42 \pm 0.57$ & $520 \pm 2.5$  & (0, 100) & (300, 800) \\
    \textcolor{black}{Radiative Cooling Time} & $\tau_r$ (s) & $73 \pm 4.1$ & $140 \pm 5.3$  & (0, 300) & (0, 300) \\
    \textcolor{black}{Sound Speed} & $c_s$ (km/s) & $7.3 \pm 0.078$ & $7.1 \pm 0.041$  & (3, 8) & (3, 8)\\
    \textcolor{black}{Horizontal Wavenumber} & $k_x$ (1/Mm) & $0.51 \pm 0.031$ & $0.38 \pm 0.11$  & (0, 1) & (0, 1) \\
    \textcolor{black}{Dispersion-Free Wave Travel Time} & $\Delta z/ c_s$ (s) & $5.7 \pm 0.061$ & $73 \pm  0.48$  & NA & NA\\
    \textcolor{black}{Acoustic Cutoff Frequency} & $\nu_{ac}$ (mHz) & $4.9 \pm 0.053$ & $5.1 \pm 0.029$  & NA & NA \\
    \hline
    \textcolor{black}{Reduced Chi-Squared} & $\chi^2_r$ & $16.79$ & $36.24$  & NA & NA \\
    \hline
    \end{tabular}
    \caption{The parameters calculated from the fits of the One-Wave model to the quiet-Sun phase travel time curves from the 2010 HMI Fe -- Fe and 2017 MOTH Na -- HMI Fe data set. These values correspond to the fits shown in Figure \ref{fig:tt-disp} (a) and (b) in black dashed lines.}
    \label{tab:owm}

    \begin{tabular}{l|l|l|l|l|l}
    \hline
    Parameter & Parameter & \multicolumn{2}{c|}{Value $\pm$ Uncertainty} & \multicolumn{2}{c}{Bounds} \\
    \cline{3-6}
    Name & Notation (Unit) & 2010 Fe -- Fe & 2017 Na -- Fe & 2010 Fe -- Fe & 2017 Na -- Fe \\
    \hline
    \textcolor{black}{Height Difference} & $\Delta z$ (km) & $49 \pm 0.63$ & $480 \pm 2.5$ & (0, 300) & (300, 800) \\
    \textcolor{black}{Radiative Cooling Time} & $\tau_r$ (s) & $77 \pm 2.9$ & $79 \pm 2.8$ & (0, 300) & (0, 300) \\
    \textcolor{black}{Sound Speed} & $c_s$ (km/s) & $6.5 \pm 0.034$ & $6.7 \pm 0.044$ & (3, 8) & (3, 8)  \\
    \textcolor{black}{Reflection Coefficient} & $R_{atm}$ & $0.13 \pm 0.0044$ & $0.12 \pm 0.0039 $ & (0, 0.3) & (0, 0.3) \\
    \textcolor{black}{Slope of Phase Change} & $\phi_m$ (deg/mHz) & $34 \pm 1.2$& $ -5.4 \pm 2.2$ & (-100,  100) & (-100,  100)\\
    \textcolor{black}{y-Intercept of Phase Change} & $\phi_c$ (deg) & $340 \pm 8.5$ & $27 \pm 15$ & (0, 360) & (0, 360) \\
    \textcolor{black}{Horizontal Wavenumber} & $k_x$ (1/Mm) & $0.56 \pm 0.034$ & $0.53 \pm 0.042$ & (0, 1) & (0, 1) \\
    \textcolor{black}{Dispersion-Free Wave Travel Time} & $\Delta z/c_s$ (s) & $7.5 \pm 0.093$ & $72 \pm 0.25$ & NA & NA\\
    \textcolor{black}{Acoustic Cutoff Frequency} & $\nu_{ac}$ (mHz) & $5.6 \pm 0.029$ & $5.4 \pm 0.035$ & NA & NA\\
    \hline
    \textcolor{black}{Reduced Chi-Squared} & $\chi^2_r$ & $0.67$ & $2.19$ & NA & NA \\
    \hline
    \end{tabular}
    \caption{The parameters calculated from the fits of the Two-Wave model to the quiet-Sun phase travel time curves from the 2010 HMI Fe -- Fe and 2017 MOTH Na -- HMI Fe data set. These values correspond to the fits shown in Figure \ref{fig:tt-disp} (a) and (b) in red solid lines.}
    \label{tab:twm}
\end{table*}

\cite{2004ApJ...613L.185F} showed that the observed phase travel time of propagating acoustic waves between two heights in the solar atmosphere, in and around active regions, depends on the location of the magnetic canopy, defined as the region where the plasma $\beta=1$ (equipartition layer for the gas and magnetic pressure). When the magnetic canopy lies between the observing heights, the wave travel time between the heights is shortened due to wave “reflection” at the canopy (see Figure 2. in \citealp{2004ApJ...613L.185F}). Interestingly, this figure also shows that the wave travel time signal at any spatial location where the canopy is above the lower observing height should consist of two components: one from the original upward traveling wave, and the other from the reflected downward traveling wave. \cite{2019ApJ...884L...8J} analyzed the phase difference observations generated from spatial locations with $|B|<30\,$G, i.e., locations where both observing heights are below the magnetic canopy, known as "quiet Sun" observations. They speculated that a ``wiggle'' in their observed phase difference versus frequency curve is the signature of a downward traveling wave. Furthermore, \cite{2015ApJ...801...27R} show in their modeling effort that halos of increased power near the active regions are caused by downward propagating waves that are the result of reflection at the $\beta\sim1$ region. Additionally, \cite{1997ApJ...485L..49J} detected wave reflection not only near the lower chromosphere but also from the base of the photosphere -- the CZP boundary.

Following these suggestions, we utilize a low-Q Fabry–Pérot interferometer \citep{1993ApJ...410..829D} wherein two consecutive reflections are considered: one from a reflective layer above the upper observing height and the other from the CZP boundary. Therefore, modifying Equation \ref{eq:v1} yields an updated velocity signal given as,

\begin{equation}
    \label{eq:v2}
    \begin{aligned}
    V(z) = V(0)e^{\kappa z}e^{i\omega t} [& e^{- i k_z z}\\
    + & R_{atm}e^{i\phi_{atm}(\omega)} e^{i k_z z} \\
 + & R_{ph}e^{i\phi_{ph}(\omega)} R_{atm}e^{i\phi_{atm}(\omega)} e^{-i k_z z} ]
    \end{aligned}
\end{equation}

where $R_{atm}\,(<1)$ and $R_{ph}\,(<1)$ are the reflectivity of wave energy at the upper atmospheric reflecting boundary and the photosphere, respectively. $\phi_{atm}(\omega)$ and $\phi_{ph}(\omega)$ are the phase shifts upon reflection from the atmospheric and photospheric reflective boundaries. Based on wave reflection analysis done using the potential barrier models in \cite{1991MNRAS.251..427W}, we assume linear dependence of the phase change on reflection for each reflective region given by $\phi(\omega) = \phi_m \omega + \phi_c$, where $\phi_m$ is the slope and $\phi_c$ is the y-intercept.

Here we study the impact of both the second and third terms in Equation \ref{eq:v2} on the quality of the fit to the data and the values of the model parameters over that produced using the first term only (One-Wave model; the traditional approach). First, we include only the atmospheric reflection (second term) and later we study the impact of the addition of photospheric reflection (third term).

The phase difference between heights $z_1$ and $z_2$, while only considering atmospheric reflection, can be calculated by inserting Equation \ref{eq:v2} in Equation \ref{eq:phi2} giving

\begin{equation}
\label{eq:dphi2}
\begin{aligned}
\Delta\Phi = &\operatorname{arg}\bigg(V(0)e^{\kappa z_1}e^{i\omega t}\big[e^{- ik_{z} z_1} + R_{atm}e^{i\phi(\omega)} e^{i k_{z} z_1}\big] \\
\times &V(0)e^{\kappa z_2}e^{-i\omega t}\big[e^{i k_{z} z_2} + R_{atm}e^{-i\phi(\omega)} e^{-i k_{z} z_2)}\big]\bigg).
\end{aligned}
\end{equation}

Using the additivity property of the argument function, $\operatorname{arg}(ab)=\operatorname{arg}(a)+\operatorname{arg}(b)$, definition of the $\operatorname{arg}$ operator $\operatorname{arg}(a) = \arctan\left(\frac{\operatorname{Imaginiary}(a)}{\operatorname{Real}(a)}\right)$, and that argument of a real number is zero, Equation \ref{eq:dphi2} can be simplified as:

\begin{equation}
    \label{eq:dphi3}
    \begin{aligned}
        \Delta\Phi = &\operatorname{arg}\big( e^{-ik_{z} z_1} + R_{atm}e^{i\phi(\omega)} e^{i k_{z} z_1}\big) \\
&+\operatorname{arg}\big( e^{ik_{z}z_2} + R_{atm}e^{-i\phi(\omega)} e^{-ik_{z} z_2}\big)\\
    =& \arctan\left\{\frac{-\sin{[k_z z_1]} + R_{atm}\sin{[\phi(\omega)+k_z z_1]}}{\cos{[k_z z_1]} + R_{atm}\cos{[\phi(\omega)+k_z z_1]}}\right\}\\
    &+ \arctan \left\{ \frac{\sin{[k_z z_2]} - R_{atm}\sin{[\phi(\omega)+k_z z_2]}}{\cos{[k_z z_2]} + R_{atm}\cos{[\phi(\omega)+k_z z_2]}}\right\}.
    \end{aligned}
\end{equation}

All the variables used above in Equation \ref{eq:dphi3} are discussed in Section \ref{sec:owm}. The phase travel time can be calculated from the phase difference formulation shown in Equation \ref{eq:dphi3} using Equation \ref{eq:tt}. This is now the updated wave propagation model in the solar atmosphere, which hereafter is referred to as the Two-Wave model. The Two-Wave model, which includes a downward propagating wave in addition to the One-Wave model, yields a better fit of the observed phase travel time curve than the One-Wave model. However, the correlation coefficients obtained from the fitting routine \texttt{curve\_fit} from the Python package \textit{scipy} \citep{2020SciPy-NMeth} found a 100\% correlation between $z_1$ and $z_2$. This suggests that one of these parameters can be fixed while the other remains a free parameter. We therfore fix the lower observing height, $z_2$, at $140\,$km \citep{2011SoPh..271...27F, 2014SoPh..289.3457N} as the approximate formation height for Fe I in both the 2010 HMI Fe -- Fe and 2017 MOTH Na -- HMI Fe data set. Following this, we can express the upper observing height as $z_1 = z_2 + \Delta z$, thus substituting the free parameter $z_1$ with the height difference $\Delta z$. This fixing of one parameter did not reduce the fit quality.

The Two-Wave model fits, after fixing $z_2$, are shown in Figure \ref{fig:tt-disp} (a) and (b), plot using solid red lines and the parameters corresponding to the fits are listed in Table \ref{tab:twm}. The reduced chi-squared value of the Two-Wave Model fit is \textcolor{black}{$0.67$ and $2.19$} for the 2010 HMI Fe -- Fe and 2017 MOTH Na -- HMI Fe data sets respectively, which are significantly better than that for the One-Wave model fits.

However, the sinusoidal residuals from the Two-Wave model fit suggests that our model does not reproduce the observed phase travel time completely. Therefore, we now include the photospheric reflection in the Multi-Wave model (third term in Equation \ref{eq:v2}) in an attempt to obtain a better fit. For the wave reflectivity of the photosphere, we utilize a frequency dependent formulation,

\begin{equation}
\label{eq:vor}
R_{ph} = \frac{1}{1+\exp\left({\frac{\omega^2-\omega_{ac}^2}{\omega_0^2}}\right)}    
\end{equation}

that behaves like a potential barrier (step function) at the acoustic cutoff frequency ($\omega_{ac}$) with a finite barrier width ($\omega_0$) \citep{1998MNRAS.298..464V}. The fit and residuals are shown in Figure \ref{fig:tt-disp}. \textcolor{black}{This flattens and reduces the sinusoidal residuals seen in the Two-Wave model fits, the reduced chi-squared value lowered from 0.67 to 0.65 and 2.19 to 1.18 for the 2010 HMI Fe -- Fe and 2017 MOTH Na -- HMI Fe data sets respectively. This suggests that the addition of a third wave improves modeling of the observed phase travel time. However, the increased number of parameters poses challenges with distinguishing between near-degenerate solutions and is discussed in Section \ref{sec:near-deg}}

\subsection{Parameters from the Wave Propagation Modeling}
\label{sec:near-deg}

\begin{figure*}
\gridline{
\fig{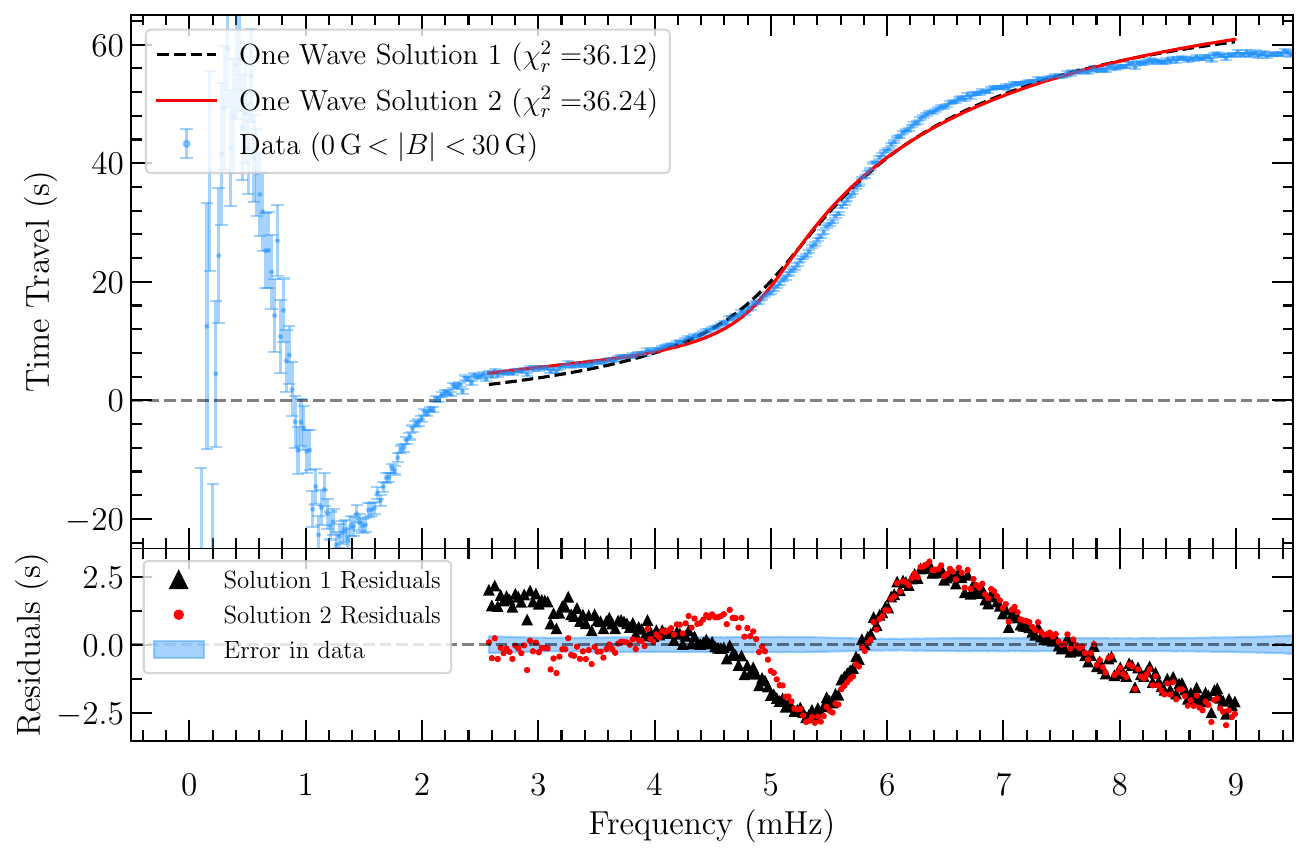}{0.5\textwidth}{(a) One-Wave Model Degeneracy}
\fig{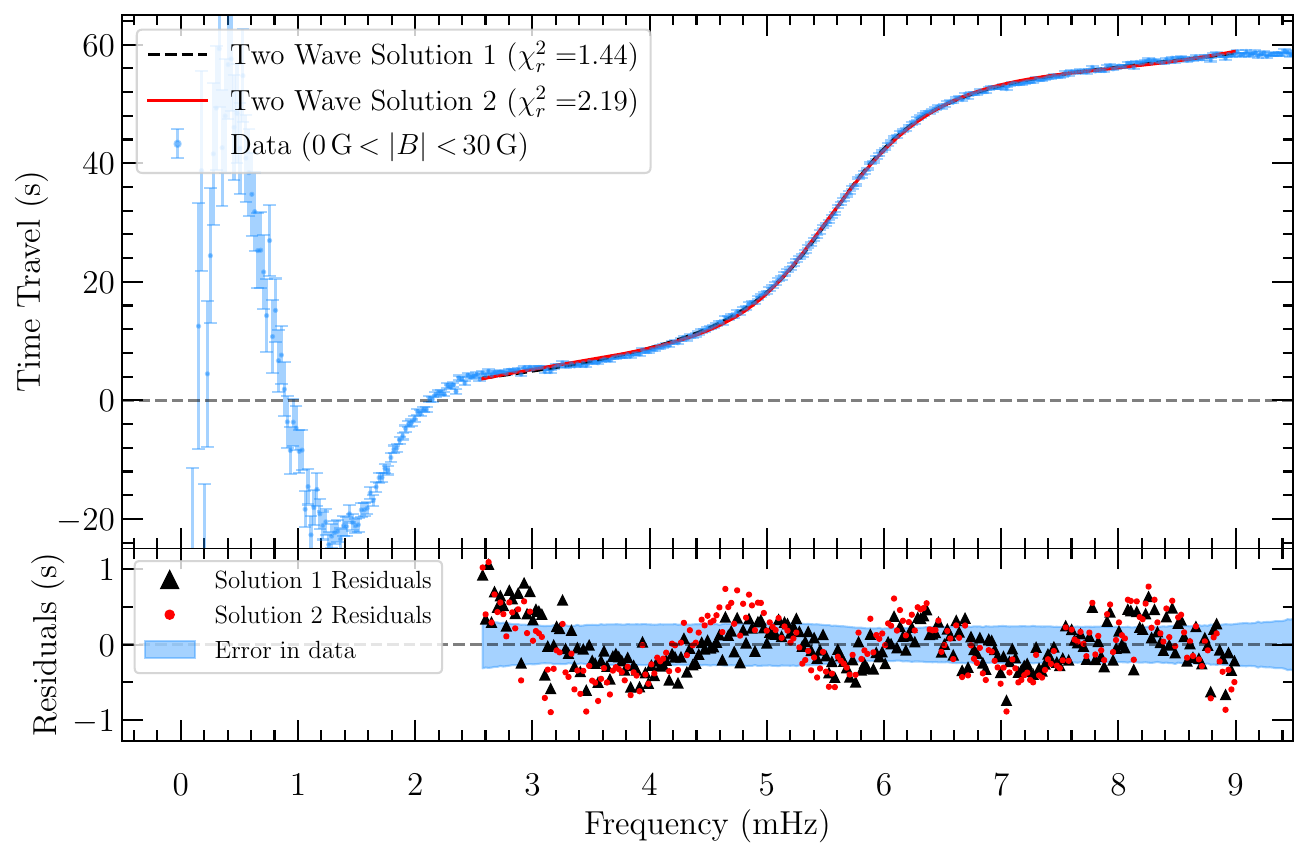}{0.5\textwidth}{(b) Two-Wave Model Degeneracy}
         }
\centering
\caption{The observed phase travel time curve for the low magnetic field region in the 2017 MOTH Na -- HMI Fe in blue filled circles with vertical error bars. (a) The black dashed lines show Solution 1 of the One-Wave fit and the black triangles plot the corresponding residuals. The red solid line shows Solution 2 of the One-Wave fit and the red filled circles show the corresponding residuals in the bottom panel. The reduced chi-squared value for Solution 1 and Solution 2 is 36.12 and  36.24 respectively. \\
One-Wave Solution 1: $\Delta z=370\,$km, $\tau_r=17\,$s, $c_s=5.7\,$km/s, $k_x=0.0\,/$Mm\\
One-Wave Solution 2: $\Delta z=520\,$km, $\tau_r=141\,$s, $c_s=7.1\,$km/s, $k_x=0.38\,/$Mm\\
(b) Same as panel (a) but for the Two-Wave model. The reduced chi-squared value for Solution 1 and Solution 2 is 1.44 and 2.19 respectively. \\
Two-Wave Solution 1 : $\Delta z=370\,$km, $\tau_r=22\,$s, $c_s=5.5\,$km/s, $R_{atm}=0.14$, $\phi_m=-31\,$deg/mHz, $\phi_c=200\,$deg, $k_x=0.0\,/$Mm\\
Two-Wave Solution 2: $\Delta z=480\,$km, $\tau_r=78\,$s, $c_s=6.7\,$km/s, $R_{atm}=0.12$, $\phi_m=-5.4\,$deg/mHz, $\phi_c=27\,$deg, $k_x=0.53\,/$Mm
\label{fig:ow-deg}
}
\end{figure*}

The One-Wave model discussed in \ref{sec:owm} was found to exhibit  degeneracy in the solution space while fitting the observed travel time data. Figure \ref{fig:ow-deg} (a) shows the 2017 MOTH Na -- HMI Fe phase travel time data with two near-degenerate One-Wave solutions. The caption lists the two solutions sets that have chi-squared values of 36.12 and 36.24. These values are within 3-standard deviations ($\sigma_{\chi^2_r}$) of the probability distribution from which the reduced chi-squared values are calculated, where $\sigma_{\chi^2_r} = \sqrt{2/N} = \sqrt{2/245} = 0.09$ \citep{2010arXiv1012.3754A} where $N$ is the total number of data points in the observed phase travel time. Hence, the two solutions are nearly identical and cannot be distinguished based on the goodness of the fit without reducing noise in the data. However, through prior understanding of the horizontal wavenumber, it is possible to eliminate one of the solutions. Based on the pixel size of the data sets used here ($\sim 1\,$Mm), a non-zero horizontal wavenumber is expected of approximately $0.5\,$/Mm. Therefore, in this case Solution 2 would be preferred.

The Two-Wave model also demonstrates a solution space where the lowest chi-squared fit solution has non-physical values of the horizontal wavenumber. Figure \ref{fig:ow-deg} (b) shows two solutions of the Two-Wave model for the 2017 MOTH Na -- HMI Fe phase travel time curve. The Two-Wave Solution 1 is discarded due to its null horizontal wavenumber, despite the low chi-squared value and Solution 2 is preferred because it provides a non-zero horizontal wavenumber, irrespective of its lower chi-squared value.

The common model parameters obtained from the One-Wave and Two-Wave model fits, as shown in Tables \ref{tab:owm} and \ref{tab:twm}, exhibit fundamental differences. Between the two models, the height difference and sound speed differ by more than three times the standard errors. The radiative cooling time for the 2017 Na -- Fe data set also differs across the two models but is essentially identical for the 2010 Fe -- Fe data set. This emphasizes the fact that the errors in the model parameters are small enough to characterize the Two-Wave model estimates of the properties of the solar photosphere as fundamentally different from the One-Wave model.

The solution space for the Three-Wave model has multiple near-degenerate parameters. These near-degenerate solutions are within 3$\sigma_{\chi^2_r}$ of the probability distribution from which the reduced chi-squared values are calculated. Given the multiple degenerate solutions to the Three-Wave model, it is challenging to identify the correct solution set and hence the parameters are not reported here.

The solutions for all wave propagation models discussed above were found via Monte Carlo simulations. We utilized 10,000 random initial parameter guesses between the bounds mentioned in Table \ref{tab:owm} and \ref{tab:twm} and allowed them each to individually fit the observed phase travel time data. Following this we rank the solutions by their reduced chi-squared values and analyze those closest to unity. The two solutions listed for each of the One-Wave and Two-Wave solutions are the top two solutions found using the Monte-Carlo simulations.

\subsection{Wave Reflection Characteristics in Magnetic Field Regions}
\label{sec:worrall}

\begin{figure*}
\gridline{
\fig{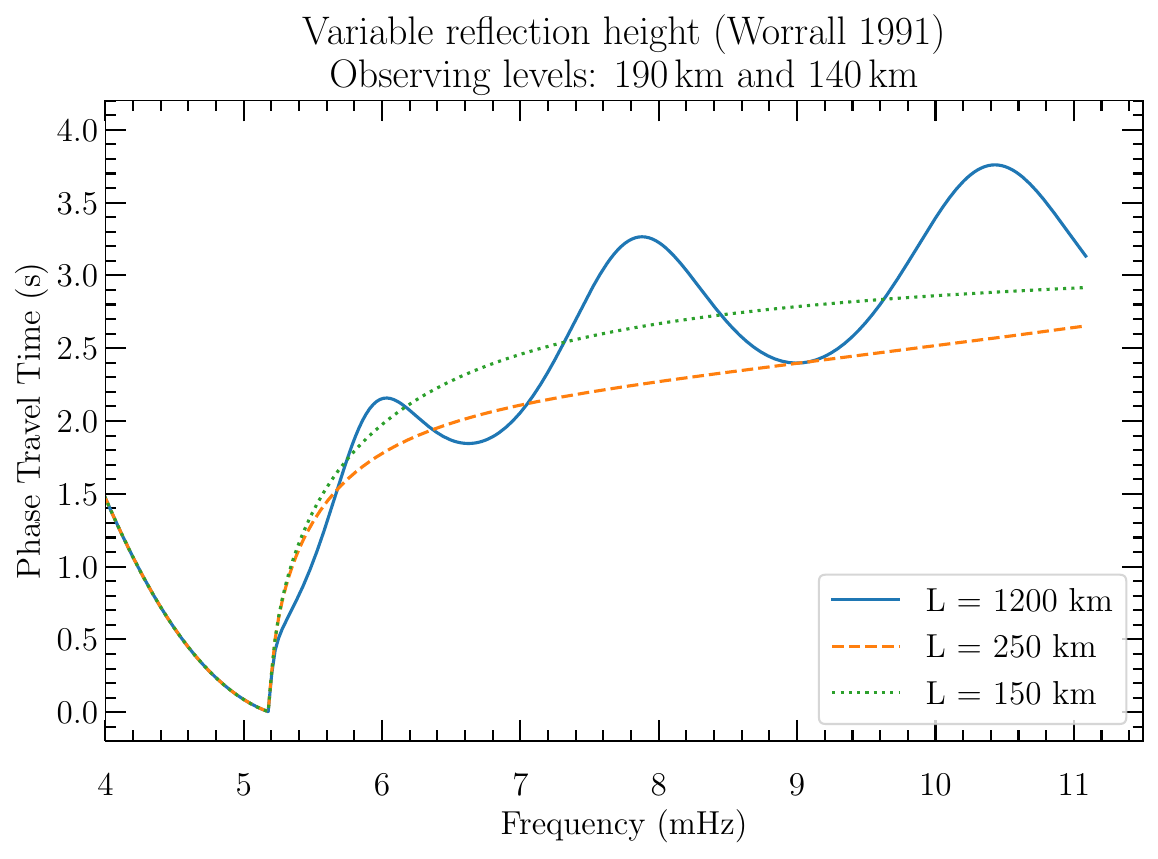}{0.5\textwidth}{(a) Analogous to 2010 HMI Fe -- Fe data set}
\fig{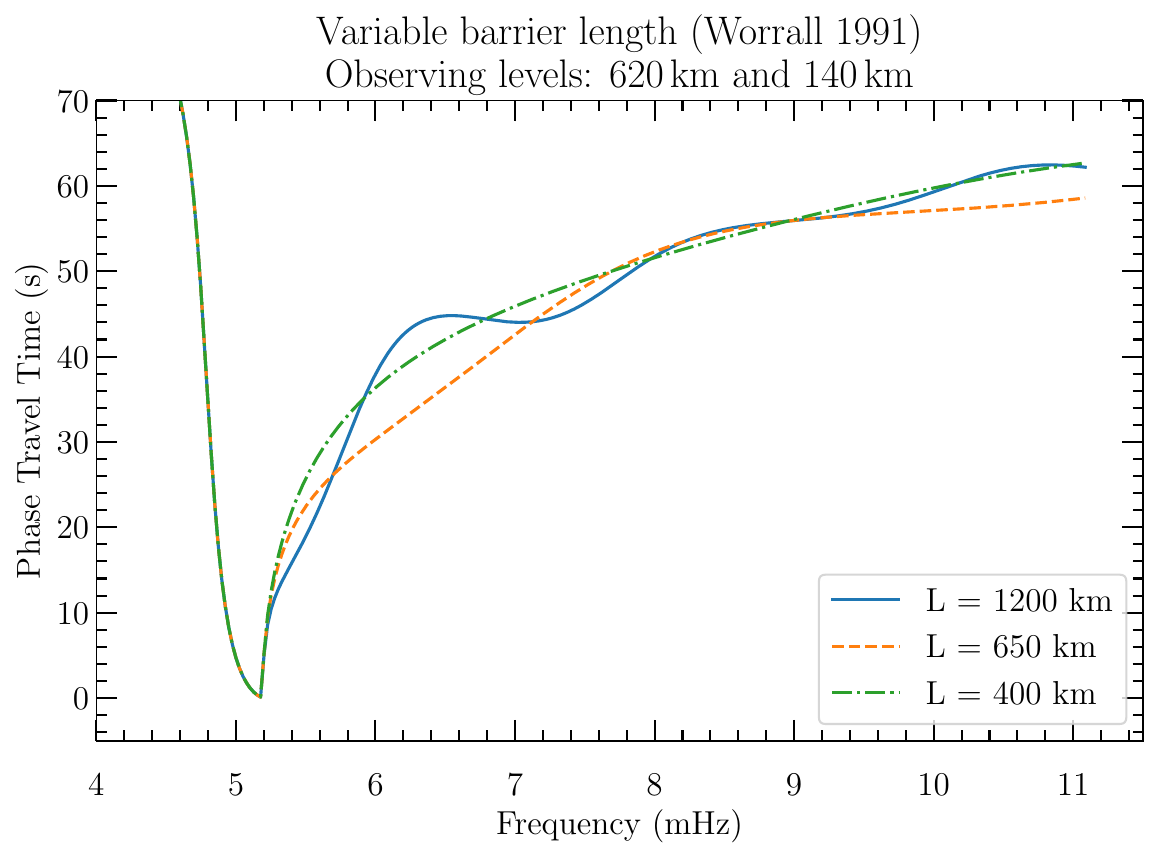}{0.5\textwidth}{(b) Analogous to 2017 MOTH Na -- HMI Fe data set}
         }
\centering
\caption{Simulated phase travel time curves for varying heights of the reflective surface model using a rectangular barrier model \cite{1991MNRAS.251..427W}. Panels (a) and (b) show two different sets of heights of in the atmosphere analogous to the 2010 HMI Fe -- Fe and 2017 MOTH Na -- HMI Fe data set respectively. The height of the reflective surface (L) corresponding to each of the phase travel time curves is shown in the legend. It is noteworthy that the ``wiggle'' in the curves disappears as the height of the reflective surface lowers closer to the upper observing level and completely disappears when the reflective height is below the upper observing height. \label{fig:worrall} }
\end{figure*}

The effects of magnetic fields on the observed phase travel time curve is explored here. The wave reflection feature present in the phase travel time curve of the quiet-Sun region gradually disappears as we move to intermediate and high-magnetic field strengths as can be seen in Figure \ref{fig:tt-disp3}. We reproduce this behavior using a rectangular potential barrier model for wave propagation to understand the driving mechanisms behind the response of phase travel time curves to magnetic fields.

In the context of acoustic waves in the Sun, the variation of the acoustic cutoff frequency with height in solar atmosphere can be approximated as a rectangular barrier (refer Figure 1 in \citealp{1991MNRAS.251..427W}). The rectangular model can be used to simulate the propagation of acoustic waves in the solar atmosphere and measure the phase travel time between two heights. The length of the rectangular barrier can be varied, which is analogous to changing the height of the reflection surface in the low chromosphere, that has been assumed to be the source of downward propagating waves in Section \ref{sec:twm}.

We simulated phase travel time curves in the solar atmosphere using a rectangular potential barrier model \citep{1991MNRAS.251..427W} and probing the atmosphere at two sets of heights. The partial reflection at the backside of the potential barrier is used to simulate the reflective layer in the atmosphere. Figure \ref{fig:worrall} (a) and (b) show the simulated phase travel time curve for 190\,km – 140\,km and 620\,km – 140\,km height combinations respectively for varying barrier lengths (or reflection heights). These height combinations in (a) and (b) are meant to draw a parallel with the 2010 HMI Fe –- Fe and 2017 MOTH Na -- HMI Fe data sets respectively. The three barrier lengths were chosen to represent a reflective region far from the upper observing level, another slightly above the upper observing level, and finally one between the lower and upper observing levels. The heights of the reflective layer were varied to study and compare analogous effects seen in the observed phase travel time curves with varying magnetic fields. This is discussed further in Section \ref{sec:res}.


\section{Results and Discussion}
\label{sec:res}

In this work, we show that the observed phase travel time at low magnetic field strengths can be modeled with increased accuracy when a downward traveling-wave component is introduced to an only-upward traveling wave model, resulting in a Two-Wave model. 

We note that for intermediate and high magnetic field strengths we lose the ability to detect the signature of a downward traveling wave in the observed phase travel time curve (i.e. the ``wiggle'' is not detected) as can be seen in Figure \ref{fig:tt-disp3}. This is to be expected if the conjecture of \cite{2004ApJ...613L.185F}, that wave reflection is occurring at the magnetic canopy (i.e., the $\beta\sim1$ region), is correct. We test this using a potential barrier model in Section \ref{sec:worrall}. To simulate the lowering of the $\beta\sim1$ region with an increasing magnetic field, we shorten the length of the potential barrier to simulate the reflection boundary moving closer to the observing heights. As the length of the barrier shortens, the amplitude of the ``wiggle'' in the phase travel time curve also reduces as can be seen in Figure \ref{fig:worrall}. Thus, the reduced level of the reflection signature in the observed phase travel time curve for stronger magnetic fields strengths supports the idea that the magnetic canopy is the source of wave reflection. The phase travel times at intermediate and high magnetic field strengths can therefore be adequately fitted using an only-upward traveling wave model (One-Wave model), as discussed in \cite{2019ApJ...884L...8J}. 

It is worth noting that the reflection signature within the phase travel time curve is not solely determined by the height of the reflection boundary; it is also influenced by the positions of the observation levels. This becomes evident when comparing the amplitude of the reflection ``wiggles'' between the simulated phase travel time curves shown in panels (a) and (b) of Figure \ref{fig:worrall}. Either panel is associated with different heights and they exhibit varying amplitudes due to the differing locations of the observation levels relative to the reflective region placed at 1200\,km.

In addition to evidence for downward propagating waves and a reflective canopy in the lower chromosphere, the Two-Wave model offers the ability to extract the physical properties of the solar atmosphere in the quiet-Sun. These properties include the acoustic cutoff frequency, radiative cooling time, reflection coefficient, and dispersion-free wave travel time ($\Delta z/c_s$: ratio of height difference and sound speed). While the One-Wave model can also estimate all the quantities except the reflection coefficient, the poor fit quality suggests that the One-Wave model fit parameters are less reliable than those obtained from the Two-Wave model. 

The Two-Wave model provides an estimate of the reflectivity of the magnetic canopy when it is above the upper observing height, i.e., for quiet-Sun regions. We measure an $R_{atm}\sim 12-13\%$. \cite{1998MNRAS.298..464V} calculated a reflectivity of $13\%$ to $26\%$, depending on the excitation source's parity (monopole, dipole, quadruple). \cite{1994ApJ...422L..29K} place an upper limit to the wave reflectivity at $\sim10\%$ and \cite{1997ApJ...485L..49J} suggest a range of reflectivity in the atmosphere of $3-9\%$. The magnitude of wave reflection and the height at which it occurs has implications for the mechanical heating of the chromosphere by acoustic waves.

Our models also provide estimates for the radiative cooling time, which is expected to increase with height in the solar atmosphere \citep{1970A&A.....4..189S}. As the 2010 Fe – Fe data set probes two heights in the lower photosphere using the Fe I line (140 -- 200\,km; \citealp{2011SoPh..271...27F, 2014SoPh..289.3457N}) while the 2017 Na -- Fe data set probes one height each in the lower photosphere (Fe I -- 160\,km) and lower chromosphere (Na D1 -- 650\,km), then we might expect to observe a longer damping time for the 2017 Na –- Fe data set. The values from our One-Wave model, $72\pm4.1\,$s for the 2010 Fe – Fe data and $140\pm5.3\,$s for the 2017 Na –- Fe data, agree with this expectation. However, the values from our Two-Wave model suggest a cooling time of $\sim78\,$s for both data sets. That is, the radiative cooling time is essentially independent of height over the range of heights sampled by our observations. Interestingly, \cite{1989A&A...224..245F} estimate a radiative cooling time of 60\,s between Fe I 593\,nm and Na $D_1$. This value aligns more closely with the estimate from our Two-Wave model than from the One-Wave model and the predictions of \cite{1970A&A.....4..189S}. 

The sound speed calculated using the Two-Wave model is \textcolor{black}{$6.5-6.7\,$km/s which is in close agreement with previously reported values of $6.8\,$km/s \citep{1989A&A...224..245F} and $7.07\,$km/s} \citep{2012SoPh..279...43W}. The sound speed estimated using the One-Wave model is  $7.1-7.3\,$km/s and is in close agreement with the Two-Wave model. The quiet-Sun wave travel time $(\Delta z/c_s)$, measured using the Two-Wave model is \textcolor{black}{$7.5\,$s} for the 2010 HMI Fe -- Fe data set and \textcolor{black}{$72\,$s} for the 2017 MOTH Na -- HMI Fe data set. This is the dispersion-free wave travel time between the lower observing layer and the $\beta\sim1$ canopy if this surface crosses between the two observing heights, or the travel time between the heights otherwise \citep{2004ApJ...613L.185F}. \textcolor{black}{These travel times correspond to a height difference of $49\,$km and $480\,$km} for the 2010 Fe -- Fe and 2017 Na -- Fe data sets respectively. \cite{1989A&A...224..245F} measured a height difference of $440\,$km between Na $D_1$ and  Fe I 593.0 nm spectral line formation heights. The 2017 Na -- Fe data set used here contains Doppler data from MOTH Na $D_1$ and HMI Fe 617.3 nm spectral lines. The Fe 617.3 nm spectra line is formed below the Fe I 593.0 nm line, thus explaining a larger height difference of $480\,$km. As for the 2010 HMI Fe -- Fe separation between the HMI 1st and 2nd Fourier coefficients, \cite{2014SoPh..289.3457N} suggests a maximum separation of $70\,$km which agrees with our measured height difference of $49\,$km. It should be noted that the sound speed and height difference present a correlation coefficient of 88\%. \cite{2019ApJ...884L...8J} emphasize that the solar atmosphere is highly corrugated and the line contribution functions vary considerably (by several hundred kilometers) over the solar surface (see, e.g.,
Figures 2 and 3 in \citealp{2011SoPh..271...27F}). Therefore, the height measurements made here could be affected by the instruments and techniques used to sample line profiles.

The acoustic cutoff frequency, measured as a function of the sound speed ($\nu_{ac}=\gamma g /4\pi c_s$), was estimated to be 5.4 -- 5.6\,mHz. These values are commensurate with the expected acoustic cutoff frequency values in the solar photosphere of $5-6\,$mHz estimated by \cite{2020A&A...640A...4F}. The One-Wave model estimates an acoustic cutoff frequency of 4.9 -- 5.1\,mHz which is lower than that measured by the Two-Wave model but is in agreement with the lower range of 4.50 -- 5.14\,mHz suggested by \cite{2002MNRAS.335..628W}. The reason for the difference between the One-Wave and Two-Wave model estimates could be attributed to the lack of flexibility in the former to distinguish between the effect of wave reflection and acoustic cutoff frequency on the phase travel time curve.

\textcolor{black}{While the One-Wave and Two-Wave models provide interesting insights into some of the physical properties of the photosphere, the models exhibit near-degeneracy. To combat this, the fitting routines were assisted with bounds to the parameters based on prior works. As discussed in this section above, the formation heights, sound speed, and atmospheric reflection coefficient have been extensively discussed in prior works. We use these as a guide to set the bounds to parameters in our fitting routines. In addition to these bounds, another distinguishing feature between degenerate solutions was the measurements of the horizontal wavenumber. Based on the pixel size of the data sets used here ($\sim 1\,$Mm), a non-zero horizontal wavenumber of about $\sim 0.5\,$/Mm is expected. However, some One-Wave and Two-Wave solutions suggested a null horizontal wavenumbers which were thence disregarded. The bounds for all the parameters are listed in Tables \ref{tab:owm} and \ref{tab:twm}. In light of this degeneracy, we recommend caution while using the One-Wave and Two-Wave model to estimate solar atmospheric properties.}

An additional reflection from the photosphere-convection zone boundary was considered in the form of a Three-Wave model that was able to flatten the residuals and provide a lower chi-square value while fitting the phase travel time data. However, the Three-Wave model has multiple near-degenerate solutions due to the high density of local minima caused by an increased number of parameters in the model and noise in the data. This limitation can be addressed by using longer data sets that will improve the frequency resolution and hence the signal-to-noise ratio of the phase travel time data.

Additionally, the linear approximation of the phase change on reflection may convey valuable insights into the height of the reflective region. The equations in the Appendix of \cite{1991MNRAS.251..427W} show that the information on the length of the barrier ($L$) is contained in the phase difference of the downward wave with respect to the incident upward wave (the phase angle $\theta$ depends on L). The height of the reflective canopy can also, in principle, be estimated using frequency-filtered cross-correlation and auto-correlation functions of the lower and upper observed heights \citep{1997ApJ...485L..49J, 2004ApJ...613L.185F}. This is, however, outside the scope of current analysis, and is recommended for future efforts undertaken to estimate the height of the reflective magnetic canopy at $\beta\sim1$.

\section{Conclusion}
\label{sec:conclusion}

In this work, we provide compelling evidence for wave reflection in the solar atmosphere. This was achieved by introducing downward propagating plane-waves in addition to upward traveling plane-waves to model wave propagation in the solar atmosphere. Our analysis of the phase travel time vs. frequency data suggests that a reflective canopy exists between the base of the photosphere and the chromosphere-corona transition region. Based on suggestions in previous works, our observed phase travel time analysis, and simple simulations of a rectangular barrier model of the solar atmosphere, we propose that a reflective canopy exists near the plasma $\beta\sim1$ region. The height of the reflective canopy is a function of the magnetic field strength and is found to move to lower heights at higher magnetic field strengths. We measure a reflectivity coefficient of $12-13\,$\% for the canopy in a magnetically quiet region of the Sun. The reflectivity of the solar atmosphere likely influences the energy budget of solar atmospheric heating via shock waves. Moreover, the presence of a reflecting layer in the lower chromosphere may provide an explanation for the chromospheric resonances reported by \cite{2002MNRAS.335..628W}.

Furthermore, upon including wave reflection in the wave propagation model, we measure an acoustic cutoff frequency of $\sim5.5\,$mHz corresponding to a sound speed of $\sim6.6\,$km/s. The radiative cooling time was measured to be $\sim78\,$s. The near identical radiative cooling times measured here for differently spaced observing levels is not expected based on theoretical estimates. As the radiative cooling time impacts the mechanical wave energy budget for chromospheric heating, it is imperative to accurately measure the wave energy lost via radiative cooling. Clearly more work needs to be done to nail down the behavior of radiative damping with height in the atmosphere.  An obvious first step here is to acquire data that sample multiple heights in the solar atmosphere, and then model the resulting travel time data for all the heights, simultaneously using a single multi-wave model that includes prescriptions for the height variation of both the radiative damping time and the sound speed.

Moreover, we measure the dispersion-free wave travel time between the lower observing level and either the upper observing level or the reflective canopy, whichever is lower. Therefore, propagating acoustic waves can be used to determine the location of the magnetic canopy at all spatial locations where the canopy is above the lower observing level and below the upper observing level. This provides a robust alternative to measuring the local magnetic field strength and using models of the solar atmosphere, to provide a measure of the local gas pressure since the magnetic canopy is found to be near the $\beta\sim1$ region.

The One-Wave model has been used for the last 40 years to derive properties of the solar atmosphere such as the height difference between spectral formation heights, acoustic cutoff frequency, and radiative cooling time. The Two-Wave model in addition not only provides an estimate of the reflection coefficient in the solar atmosphere but vastly improves the fitting quality over the One-Wave model. Although the Two-Wave model provides better estimates of all the parameters, both the One-Wave and Two-Wave models exhibit near-degenerate solutions while fitting the concerned observed phase travel time data. While the correct solution is usually embedded in a sea of local minima, they can be extracted using Monte Carlo simulations and the degeneracy broken by incorporating prior knowledge on the parameters. It is therefore essential to exercise caution while approaching the use of wave propagation models for estimating properties of the solar atmosphere.\\

\noindent
The authors extend sincere gratitude to those instrumental in fabricating and upgrading the MOTH II instruments and collecting data at the South Pole in 2016/17: Wayne Rodgers, Cynthia Giebink, William Giebink, Les Heida, Gary Nitta, and Stefano Scardigli. The South Pole mission was funded by NSF award OPP 1341755. We also appreciate NASA/SDO and the HMI science team for providing HMI data. The collective contributions of the SDO mission and, particularly, the HMI instrument team, are recognized. Special thanks to Sebastien Couvidat for sharing the crucial 2nd Fourier coefficient HMI Doppler data.


\bibliography{sample631}{}
\bibliographystyle{aasjournal}



\end{document}